%% file: main.tex
\documentclass[manuscript]{acmart}
\AtBeginDocument{%
  \providecommand\BibTeX{{%
    \normalfont B\kern-0.5em{\scshape i\kern-0.25em b}\kern-0.8em\TeX}}}

\setcopyright{acmcopyright}
\copyrightyear{2025}
\acmYear{2025}
\acmDOI{XXXXXXX.XXXXXXX}

\acmJournal{TOCHI}
\acmVolume{1}
\acmNumber{1}
\acmArticle{1}
\acmMonth{1}

\usepackage{subcaption}
\usepackage{wrapfig}
\usepackage{color}

\usepackage{amsmath}
\usepackage{booktabs}
\usepackage{multirow}
\usepackage{rotating}
\usepackage{dblfloatfix} 
\usepackage{adjustbox}
\usepackage{graphicx}
\usepackage{lscape}
\usepackage{enumerate}
\usepackage[normalem]{ulem}

\begin{document}



\title{Who Leads? Comparing Human-Centric and Model-Centric Strategies for Defining ML Target Variables}

\author{Mengtian Guo}
\affiliation{%
  \institution{University of North Carolina at Chapel Hill}
  \city{Chapel Hill}
  \country{USA}}
\email{mtguo@unc.edu}
\author{David Gotz}
\affiliation{%
  \institution{University of North Carolina at Chapel Hill}
  \city{Chapel Hill}
  \country{USA}}
\email{gotz@unc.edu}
\author{Yue Wang}
\affiliation{%
  \institution{University of North Carolina at Chapel Hill}
  \city{Chapel Hill}
  \country{USA}}
\email{wangyue@unc.edu}





\input{tex/abstract}








\maketitle

\input{tex/introduction}

\input{tex/related_work}
\input{tex/method}

\input{tex/experiment}

\input{tex/results}

\input{tex/discussion}

\input{tex/conclusion}

\begin{acks}

\end{acks}

\bibliographystyle{ACM-Reference-Format}
\bibliography{main}

\appendix









\end{document}

%% file: tex/abstract.tex
\begin{abstract}

Predictive modeling has the potential to enhance human decision-making. However, many predictive models fail in practice due to problematic problem formulation in cases where the prediction target is an abstract concept or construct and practitioners need to define an appropriate target variable as a proxy to operationalize the construct of interest. 
The choice of an appropriate proxy target variable is rarely self-evident in practice, requiring both domain knowledge and iterative data modeling. This process is inherently collaborative, involving both domain experts and data scientists. In this work, we explore how human-machine teaming can support this process by accelerating iterations while preserving human judgment. We study the impact of two human-machine teaming strategies on proxy construction: 1) relevance-first: humans leading the process by selecting relevant proxies, and 2) performance-first: machines leading the process by recommending proxies based on predictive performance. Based on a controlled user study of a proxy construction task ($N = 20$), we show that the performance-first strategy facilitated faster iterations and decision-making, but also biased users towards well-performing proxies that are misaligned with the application goal. Our study highlights the opportunities and risks of human-machine teaming in operationalizing machine learning target variables, yielding insights for future research to explore the opportunities and mitigate the risks.

\end{abstract}

%% file: tex/introduction.tex
\section{Introduction}

Predictive modeling has recently attracted much attention from organizations trying to leverage AI and machine learning (ML) to enhance work processes such as decision-making. However, many predictive AI applications fail to support decision-makers in practice, at times having the potential to cause negative social impact. While many reasons can cause such failures, existing works pointed out that using problematic prediction targets is a major reason
~\cite{passi_problem_2019, Guerdan_groundless_2023, wang2024against, coston_validity_2023}.

Choosing the right prediction target for a predictive AI application is not always straightforward.
Some prediction targets that we would like to predict are conveniently observed in training data, such as the star rating of a product review and the readmission of a discharged patient within 30 days. Other prediction targets are unobserved in data. These unobserved targets are typically abstract constructs of interest that we aim to predict, including a student’s wellbeing, a patient’s need for health care, or a job applicant’s prospect of being a good employee. To train a ML model that predicts the construct of interest, we must define a \textbf{proxy target variable}\footnote{Throughout the paper, we  use ``proxy target variable'', ``proxy target'', and ``proxy'' interchangeably.}  that is empirically measurable and closely captures the construct of interest. The process of defining a proxy target variable to measure an unobservable construct of interest is termed ``operationalization''  in research design~\cite{operationalization}. The selected proxy can significantly influence the resulting model's predictive performance and utility in real-world applications. In high-stakes scenarios such as medical treatment~\cite{ pfaff_identifying_2022}, child welfare~\cite{kawakami2022improving, vaithianathan2017developing}, hiring~\cite{tambe_artificial_2019, van2021machine}, education~\cite{liu2023reimagining}, and special financing~\cite{passi_problem_2019}, researchers have identified cases where wrongly specified proxy target variables resulted in failed projects, predictive models that are avoided by experts, or biased predictions in real-world applications~\cite{obermeyer_dissecting_2019, kawakami2022improving, van2021machine}. 

The process of selecting a proxy to operationalize a construct of interest is often described as the problem formulation or model specification stage in AI application development~\cite{niaksu_crisp_2015, fayyad_kdd_1996}. Ethnographic work focusing on data science practices has highlighted the importance of collaboration between data scientists and domain experts in this stage~\cite{mao_how_2019, zhang_how_2020, nahar_collaboration_2022, pei_requirements_2022, kross_orienting_2021}. The two roles contribute to the project with complementary expertise. While domain experts can identify proxies that are theoretically sound and aligned with task goals, data scientists can evaluate whether predicting a proxy is technically feasible given the available data and resources. 

To illustrate the real-world challenges of AI problem formulation and the collaboration needed between different roles, consider a fictional data scientist, Mike, who was collaborating with clinicians to develop a model for predicting a patient's risk of having sepsis. Here, the construct of interest was the onset of \emph{sepsis}, a life-threatening condition caused by the body's extreme response to an infection. To operationalize sepsis using measurable data in electronic health records (EHRs), the clinicians considered multiple proxies.  These included (1) ``a set of sepsis-indicating events occurring within a 6-hour window'', (2) ``a set of sepsis-indicating events occurring during a patient's entire stay'', and (3) ``patient mortality''~\cite{sendak_real-world_2020, moor2021early}. After collaborative discussions, the team agreed on the first proxy. Mike proceeded to perform extensive data wrangling to prepare training, validation, and test data, and developed a sepsis risk prediction model. As he evaluated the model, he found that it performed poorly and produced excessive false alerts due to class imbalance --- few patients in the EHR had the set of sepsis-indicating events occurring in 6 hours~\cite{lauritsen2021framing}.
Despite experimenting with various data rebalancing methods and ML models, the results remained unsatisfactory. Mike and the clinicians revisited the candidate proxy targets to explore alternatives that may reduce class imbalance and result in better performance.
Evaluating these alternatives would require Mike to reprocess the entire EHR data, rebuild the model, and re-evaluate its performance.  There was no guarantee of performance improvement since each proxy defined a different prediction task.

As illustrated by the previous example, the process of proxy target selection is not a purely technical task, but a sociotechnical one. It requires translating high-level task goals into operational definitions that are both task-relevant and technically feasible~\cite{passi_problem_2019,kross_orienting_2021}. Existing literature reveals that proxy target selection is a collaborative and iterative process—one in which domain knowledge and technical feasibility must continuously inform each other~\cite{passi_problem_2019, mao_how_2019, nahar_collaboration_2022, kross_orienting_2021}. When operationalizing proxy target variables, two goals need to be simultaneously achieved:
\begin{itemize}
    \item \textbf{{Predicting the right proxy}}: Choose a variable or a combination of variables that best measures the construct of interest. Otherwise, the ML model will have low utility no matter how accurate it is, because it predicts the wrong target. Domain knowledge plays an important role in this aspect. 
    \item \textbf{{Predicting the proxy right}}: The ML model should have a high-enough predictive performance for the proxy target variable. Otherwise, the ML model will have low utility no matter how sound the proxy is, because its predictions are usually wrong. Data scientists are often needed in this aspect, as data processing, modeling, and evaluation are often needed to assess the performance of the resulting model. 
    However, this aspect is often slow and resource-intensive, which limits the number of iterations.
\end{itemize}

\begin{figure}[t]
\centering  
\begin{subfigure}{.49\textwidth}
    \centering
    \includegraphics[width=1\textwidth]{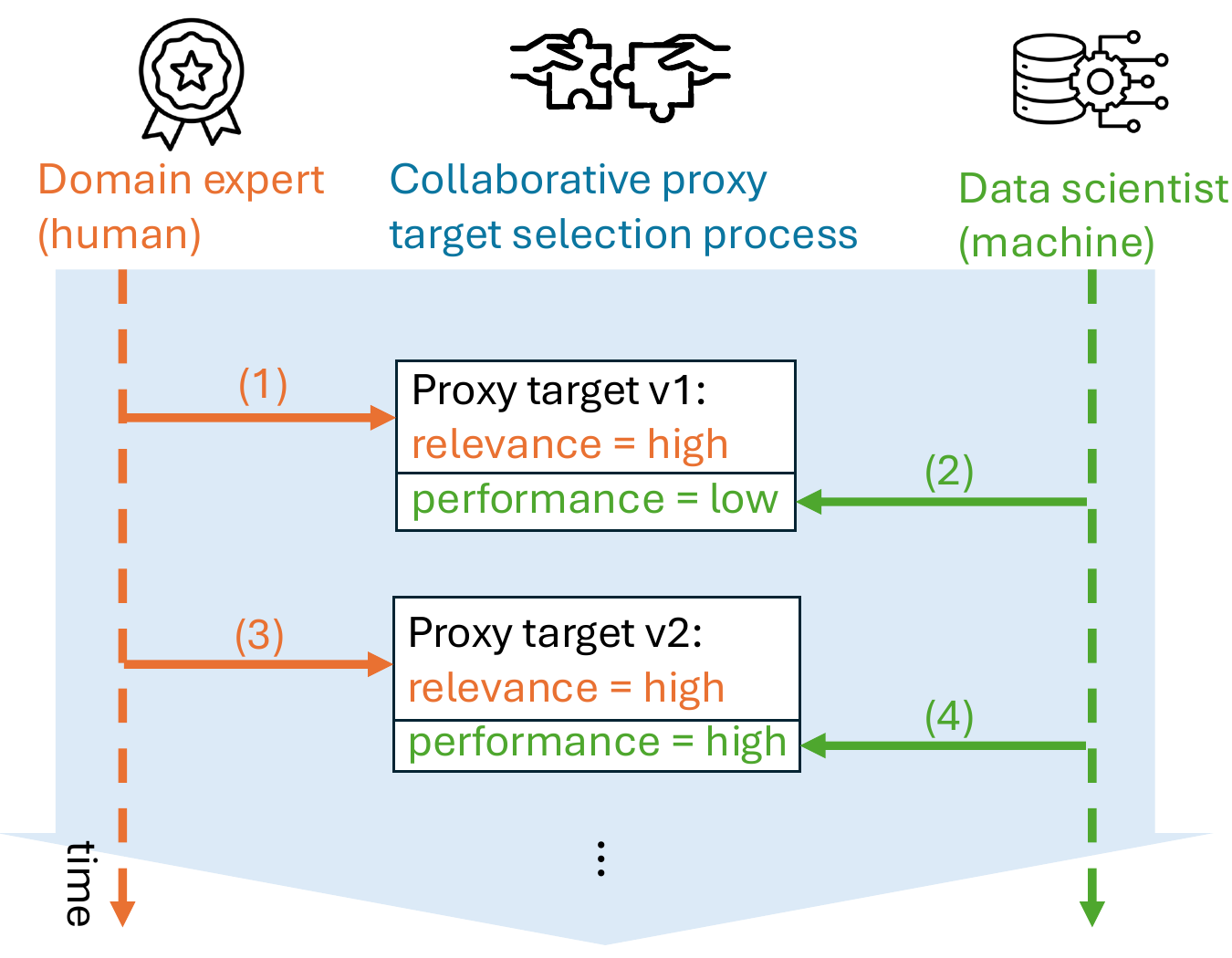}
    \caption{\textsc{Relevance First}}\label{fig:rel_first}
\end{subfigure}
    \hfill
\begin{subfigure}{.49\textwidth}
    \centering
    \includegraphics[width=1\textwidth]{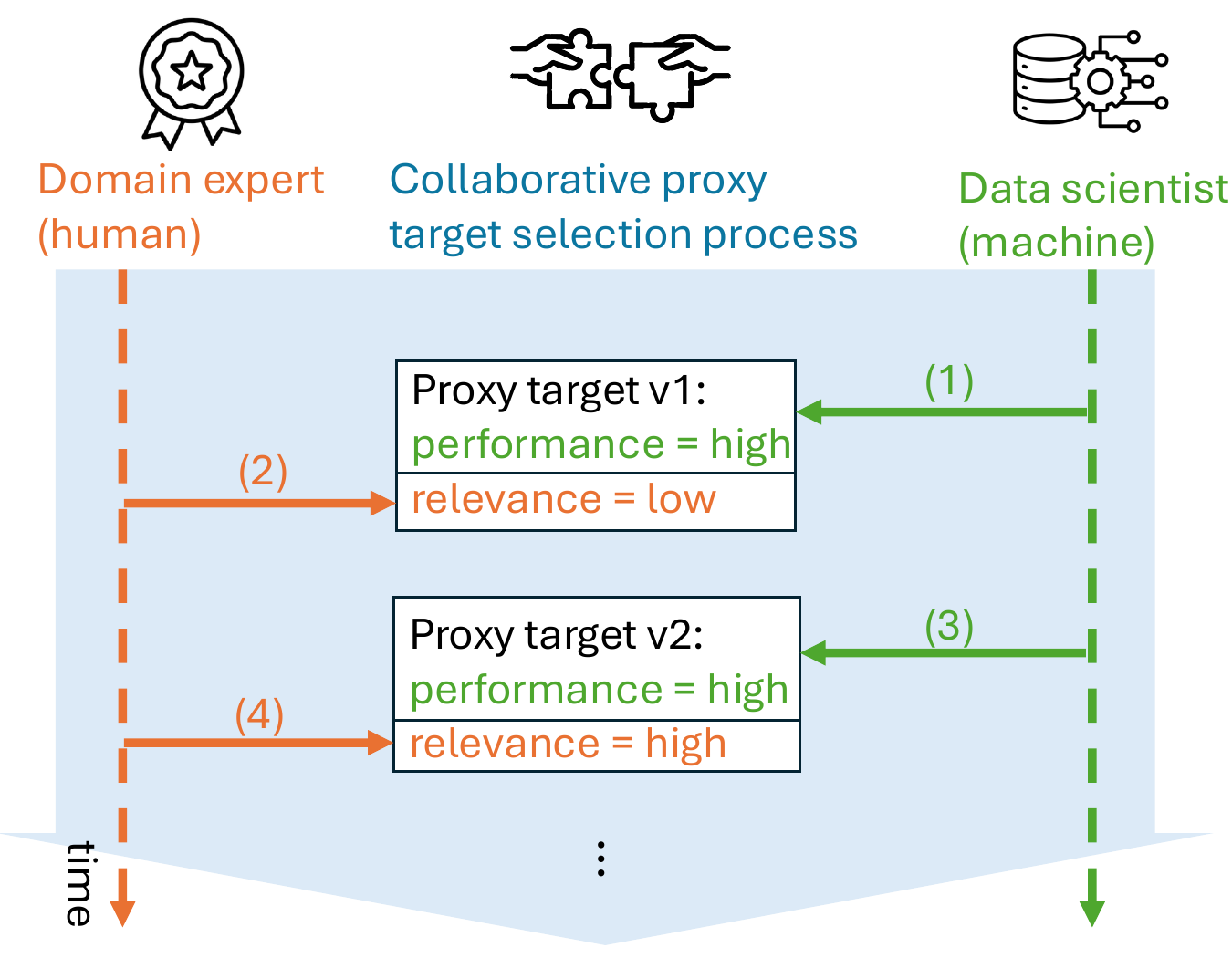}
    \caption{\textsc{Performance First}}\label{fig:perf_first}
\end{subfigure}
\caption{Illustration of two collaborative, iterative strategies for machine learning proxy target selection. Domain experts are humans with nuanced understanding of the task domain. Data scientists are humans or machines in charge of data processing, model creation, and evaluation.  In (a) \textsc{Relevance First} strategy, the domain expert proposes the next proxy with high relevance but unknown performance, then the data scientist evaluates the performance of predicting the proxy target. In (b) \textsc{Performance First} strategy, the data scientist proposes the next proxy target with high performance but unknown relevance, then the domain expert evaluates the relevance of the proxy.
}
\label{fig:workflows}
\end{figure}

In recent years, substantial efforts have been made to automate various stages of the data science process to make machine learning more accessible and efficient~\cite{he2021automl, karmaker_automl_2021, gil_towards_2019}. Automated Machine Learning (AutoML) tools have been developed to automate tasks like data preprocessing, model selection, and model evaluation~\cite{wang2021flaml, olson2016tpot}. In the problem formulation process, AutoML can serve as a rapid prototyper, enabling faster feedback in the aspect of \emph{predicting the proxy right}. However, the aspect of \emph{predicting the right proxy} still requires human input, as it is challenging to fully encode the domain knowledge and usage scenario. Inspired by the collaborative nature of the proxy selection process, we explore how to support proxy selection through a human-machine collaboration approach. Rather than fully automating proxy selection, we envision systems in which machines provide fast, iterative feedback on model feasibility, while humans remain responsible for assessing whether a proxy is meaningful and whether the resulting model is useful. 

Our work builds on a growing body of literature on data science practices~\cite{passi2018trust, kross_orienting_2021, mao_how_2019, wang_human-ai_2019, smith_enabling_2021, jung_how_2022}, the influence of automation on human decision-making~\cite{green_principles_2019, morrison_impact_2024, buccinca2021trust, chiang2021you}, and HCI research on interactive systems that support data science processes~\cite{zhang_how_2020,yang_grounding_2018, sivaraman2025tempo, cashman_userbased_2019}. While prior studies have attempted to support the collaborative and iterative process of problem formulation, they focus on a single system and qualitative evaluation, lacking empirical comparison~\cite{sivaraman2025tempo, cashman_userbased_2019}. In this paper, we endeavor to take an initial step toward understanding the effects of this human-machine team on defining the proxy target through a quantitative study. Our study shows how technologies such as AutoML can mediate the problem formulation process and providing insights for designing supportive systems.

Specifically, we considered two ways of synergizing the human-machine collaboration (Fig.~\ref{fig:workflows}):
\begin{itemize}
    \item \textsc{\textbf{Relevance First}}: a human selects a proxy target that aligns with the task goal, and then a machine reveals the predictive performance of the proxy.
    \item \textsc{\textbf{Performance First}}: a machine recommends a proxy target based on predictive performance, and then a human decides whether it aligns with the task goal.
\end{itemize}

\textsc{Relevance First} enables users to begin with proxies that are inherently relevant, reflecting a human-driven approach in which AutoML serves primarily to accelerate iterations. However, this manual process relies heavily on human intuition and may overlook proxies that are both predictable and relevant. In contrast, \textsc{Performance First} replaces manual exploration with a systematic search guided by predictive performance. Here, AutoML not only offers performance feedback but also steers the process by presenting candidate proxies to users. The trade-off is that a proxy with strong predictive performance may nonetheless lack real-world relevance. Although humans are expected to assess whether a proxy aligns with the modeling goal, their judgment may be influenced by the speed and feedback dynamics introduced through human–machine collaboration.

Through a controlled lab study, we investigate the following research questions:
\begin{itemize}
    \item[\textbf{RQ1}] How do different human-machine teaming strategies influence the quality of the final proxy target selected by a user?
    \item[\textbf{RQ2}] How do different human-machine teaming strategies influence a user’s satisfaction and decision-making experience in the proxy target selection task?
\end{itemize}

Our study provides an empirical, controlled comparison of two human-machine teaming mechanisms in a lab setting. Our findings show that while teaming with AutoML accelerates proxy selection iterations, the increased accessibility of performance feedback can bias users toward high-performing proxies, even when those proxies are less aligned with task goals or domain needs. We conclude with design recommendations for AI-assisted systems that aim to support human judgment in collaborative predictive modeling.

%% file: tex/related_work.tex
\section{Related Work}


\subsection{Challenges of Proxy Target Selection in Machine Learning Workflows}

Proxy target selection is one of the tasks during the problem formulation stage, which is the first stage of a ML workflow. In one of the most widely used ML workflows, the Cross-Industry Standard Process for Data Mining (CRISP-DM),  ML development activities are organized into six phases~\cite{wirth_crisp-dm_nodate}. The first phase, business understanding, is defined as the process of ``understanding the project objectives and requirements from a business perspective, and then converting this knowledge into a data mining problem definition.'' In CRISP-DM, this phase involves iterative refinement that is informed by later phases such as data understanding and evaluation. 


The challenge of proxy target selection has been illustrated by case studies and contextual inquiries~\cite{yang_re-examining_2020, nahar_collaboration_2022}. Below we review and summarize different aspects of this challenge.



\textbf{Understanding ML and data constraints:}
It often takes non-technical roles such as designers, domain experts, and product personnel significant efforts in understanding what ML can and cannot do~\cite{yang_re-examining_2020, nahar_collaboration_2022, yang_grounding_2018, yang_sketching_2019}. A realistic understanding of ML capabilities is essential for these roles to define feasible ML functions.
Technical roles such as data scientists often spend significant efforts in the capture, curation, design, and creation of data for ML development~\cite{feinberg_design_2017, muller_how_2019, kandel_enterprise_2012, muller_designing_2021}. Since data availability and quality impact ML problem formulation, a deep understanding of the data is essential for these roles to implement solutions that fit the application context~\cite{kandel_enterprise_2012}.

\textit{Case 1:} Lauritsen et al. \cite{lauritsen2021framing} provides a case study of different sepsis prediction problem formulations. Four commonly used problem formulations were compared. The study shows that different formulations led to a large variation in class imbalance from 1:15 to 1:750. The resulting model performance, measured by the area under precision-recall curve (AUPRC), varied greatly. For instance, the \textit{sliding window} formulation splits the entire admission into chunks, and each chunk is labeled sepsis-positive if there is an onset of sepsis. This framing allows the model to be used for sepsis risk monitoring starting from admission. However, the high frequency of prediction and the low model performance reduce its clinical utility. An alternative framing is to train the model to predict sepsis when early warning score (EWS) assessments are performed by clinical staff. This framing reduces the class imbalance in data and the frequency of model usage. 

\textit{Case 2:} Passi and Barocas \cite{passi_problem_2019} provide a case study of defining the proxy outcome variables in car financing. When developing a predictive model to match borrowers with auto dealers, the team initially identified the outcome variable to be the dealer's decision, i.e., predicting whether a dealer would approve a financing request. However, since the team did not have full access to all dealers' decisions, they had to turn to a different outcome variable that was more available. Based on the analyst's knowledge, credit score plays an important role in the special financing approval process. Credit scores thus became a proxy for a dealer’s decision and served as the outcome variable. This case study shows the trade-offs between the business objective and data availability.


\textbf{Selecting the right proxy:} Proxy targets that are misaligned with the actual goal of the application can cause unexpected outcomes and biases~\cite{passi_problem_2019, martin_jr_participatory_2020}. In application fields such as medical treatment~\cite{obermeyer_dissecting_2019}, child welfare~\cite{kawakami2022improving, vaithianathan2017developing}, education~\cite{liu2023reimagining}, hiring~\cite{tambe_artificial_2019, van2021machine}, and criminal justice~\cite{bao_its_2022}, domain experts have reported a discrepancy between the model prediction and their priority. In a lot of cases, despite high performance, the suggestions given by those predictive models were ignored by experts or were later found to have serious ethical issues.

Recent work proposed normative perspectives on ML problem formulation~\cite{coston_validity_2023, raji_fallacy_2022}. For example, Coston et al. employed modern validity theory from social science and summarized the threats to ML problem validity into three categories: attribute misalignment (including features without a plausible causal path between features and targets), target misalignment (inappropriate proxy target variables), and population misalignment (mismatch between the target population and the training data)~\cite{coston_validity_2023}. Guerdan et al.~\cite{Guerdan_groundless_2023} proposed a framework that summarizes different sources of proxy validity issues, e.g., measurement error, intervention effects, and selection bias. Informed by statistics and quantitative social sciences methods, the authors suggest that problem validity can be evaluated through lenses such as construct reliability, construct validity, outcome cross-validity. Despite using quantitative analysis, the recommended methods all involve domain knowledge and subjective human judgments.

\textit{Case 3:} Kawakami et al.~\cite{kawakami2022improving} investigated how child welfare workers reacted to the algorithms developed to assist the review of referrals for potential child maltreatment. A screening algorithm (Allegheny Family Screening Tool, or AFST) was developed to predict the  risk of child maltreatment. This goal was operationalized as predicting two outcome variables: placement in foster care and referral after screening out~\cite{vaithianathan2017developing}. However, child welfare workers reflected that these outcomes represent cases where severe maltreatment has long happened, which go against their priority --- to identify immediate risks to children. This misalignment between the proxy target  and the workers' goal led to the rejection of tools in cases where they see misaligned priorities.

\textbf{Iterative prototyping and testing:} The joint requirement of predictive performance and aligning with the modeling goal makes ML problem formulations an iterative negotiation between different roles in the ML development team. As Passi and Jackson described, data scientists ``continuously straddle the competing demands of formal abstraction and empirical contingency''~\cite{passi2018trust}. To make useful ML applications, data scientists must not just translate broader objectives into abstract modeling problems, but also negotiate these translations with non-data-scientist stakeholders \cite{passi_problem_2019}.

The iterative nature of problem formulation is repeatedly recognized in ML development workflows, including KDD~\cite{fayyad_kdd_1996}, TDSP~\cite{noauthor_team_nodate}, and CRISPML(Q)~\cite{studer_towards_2021}. Prototyping and testing help people realize the limitations of the current proxy target, leading to the reselection of the proxy target. However, due to the complexity of data and uncertainty of ML outcomes, ML prototyping is time-consuming~\cite{karmaker_automl_2021, passi_problem_2019, mao_how_2019}, often involving a labor-intensive inner-loop of ML development (i.e., data preprocessing, model training, hyperparameter optimization, and model selection). The prototyping process slows down the feedback loop, which in turn slows down proxy target selection.

\textbf{Collaboration between different roles:} Formulating the ML problem often involves people with different knowledge and expertise and it can be challenging to coordinate this collaboration~\cite{mao_how_2019, zhang_how_2020, nahar_collaboration_2022, pei_requirements_2022, kross_orienting_2021}. Technical members of the ML team often need to help non-technical members understand ML and data constraints, while non-technical members often need to provide the data and help technical members understand the data and modeling goals~\cite{nahar_collaboration_2022, kross_orienting_2021, sendak_real-world_2020, kerrigan_survey_2021}. 
Even the two groups can establish a ``common ground'' at the outset of problem formulation, synchronizing the common ground is difficult as the technical roles update their understanding of ML and data constraints and the non-technical roles update the ``right question to ask''~\cite{mao_how_2019}. The process of translating an application problem into an ML problem often appears opaque to non-technical roles, causing communication breakdowns~\cite{kross_orienting_2021, mao_how_2019}.

\subsection{Existing Tools and Systems Supporting Proxy Target Selection}

While there is a plethora of tools and systems that support developing and refining ML models, most of them are not dedicated to supporting proxy target selection. A number of human-guided machine learning systems support rapid construction of machine learning pipelines through an interactive no-code (or low-code) interface~\cite{patel_gestalt_2010, santos_visus_2019, carney_teachable_2020, honaker_statistical_nodate, du_rapsai_2023}. These systems require the user to specify a proxy target at the beginning of the process. They facilitate the experiments on different proxy targets by streamlining the ML inner loop. 

Sivaraman et al.~\cite{sivaraman2025tempo} proposed Tempo, an interactive system that helps data scientists and domain experts collaboratively iterate on model specifications. The system allows data scientists to do quick prototyping with a temporal query language and allows domain experts to assess performance within data subgroups to validate that models behave as expected. Tempo was designed to enable faster iteration based on both the model performance and the problem's alignment to the modeling goal. The system requires the users to specify problem formulations to explore, thus being similar to the \textsc{Relevance First} strategy in this paper. Our work is different in that we aim to study the usage of AutoML to streamline the prototyping process, and we provided a quantitative evaluation of the effect of faster iterations enabled by such techniques. 

Cashman et al.~\cite{cashman_userbased_2019} proposed the Exploratory Model Analysis (EMA) system, which explicitly facilitates the user to discover meaningful problems to solve on a given dataset. EMA automatically experiments on all potential proxy targets in a dataset. Different proxies and the resulting modeling performance are presented to the user to support proxy selection. This system implicitly influences users' proxy selection through predictive performance, thus similar to our study's \textsc{Performance First} strategy. Similar to the Tempo study~\cite{sivaraman2025tempo}, the EMA was evaluated qualitatively through case studies, whereas our work provided quantitative evaluation.

In summary, among the rare studies that look into strategies to support proxy target selection, none of them provided a quantitative comparison between different strategies. Our study helps to fill this gap by designing a controlled lab study and providing a quantitative comparison between different conditions.

\subsection{Tools and Studies on Multi-Attribute Choice Tasks}
Proxy target selection is a multi-attribute choice task where people need to select the best alternative among a fixed set of alternatives considering multiple attributes (e.g., task relevance and predictive performance). Multi-attribute choice tasks are closely related to Multi-Criteria Decision Making (MCDM), a field that studies procedures to aid decision making in areas like business intelligence and finance \cite{triantaphyllou2000multi}. Unlike MCDM approaches that focus on making the best selections based on user-defined criteria, our study focuses on general human-machine teaming strategies to support proxy target selection.

Several tools have been proposed to assist multi-attribute choice tasks, including domain-specific tools \cite{riehmann2012product, pu2000enriching} and general-purpose tools~\cite{elmqvist2008rolling, carenini_valuecharts_2004, gratzl2013lineup}. These tools facilitate decision-making by allowing users to iteratively refine the filtering criteria or aggregate multiple objectives to generate an overall ranking. However, in proxy target selection, the relevance attribute is difficult to quantify and often relies on human judgments, making it difficult to filter and rank considering both relevance and performance. Therefore, we focus on studying two simpler strategies that prioritize either relevance or performance to guide the selection of a proxy target.

%% file: tex/method.tex
\section{Problem Formulation for Proxy Target Selection}
\label{sec:formulation}
In this section, we formulate the proxy target selection problem and related concepts.

\textbf{Definition 1 (Target Outcome)}: The target outcome $Y^*$ is a construct of interest that the machine learning model aims to predict in an application task. It is conceived by people but unobserved in the task-specific dataset. For example, in the task of long COVID status prediction, $Y^*$ is a random variable that takes a binary value: ``having long COVID'' or ``not having long COVID''~\cite{pfaff_identifying_2022}. Although ``long COVID'' has long been conceived and discussed as a serious medical condition since 2020, the condition does not correspond to a designated clinical term or data column in electronic health records.

\textbf{Definition 2 (Observed Outcomes)}: The observed outcomes $\mathcal U = \{U_1, U_2, \cdots, U_m\}$ are a set of observed or measurable variables pre-identified in the task-specific dataset. Each observed outcome $U \in \mathcal U$ is plausibly related to the target outcome $Y^*$.  Continuing the example of long COVID status prediction, $\mathcal U$ may contain variables such as  $U_1$ = ``ICD code U09.9 is present in the patient's record,'' $U_2$ = ``self-reported symptoms are present after 4 weeks of positive COVID-19 diagnosis,'' and $U_3$ = ``self-reported symptoms are present after 3 months of positive COVID-19 diagnosis.''

\textbf{Definition 3 (Proxy Target)}: A proxy target $Y$ is a function of observed outcomes $\mathcal U$: $Y = g(\mathcal U)$. By construction, the proxy target $Y$ is also observed and measurable using the task-specific dataset. The function $g(\cdot)$ defines a syntax to select, transform, and combine one or more observed outcomes in $\mathcal U$ to construct a variable $Y$ to be used as a surrogate of the target outcome $Y^*$ based on domain knowledge. Continuing the example of long COVID status prediction, if $Y = U_1 \lor U_2$ (a Boolean syntax), then we assign a proxy label ``having long COVID'' to patients whose records had either ICD code U09.9  or self-reported symptoms after 4 weeks of positive COVID-19 diagnosis (or both), and a proxy label ``not having long COVID'' to other patients. Apparently, it takes clinical domain knowledge to construct a sensible proxy label in this context.

\textbf{Definition 4 (Predictors)}: Predictors $\mathcal X$ are observed variables pre-identified in the task-specific dataset that can be used as features in a machine learning model $f$ to predict the proxy target: $Y \leftarrow f(\mathcal X)$. Continuing the example of long COVID status prediction, predictors can include a patient's symptoms and medication during acute COVID-19 infection, their demographic information, and COVID-19 treatment measures.

Given the concepts defined above, the problem of \textbf{Proxy Target Selection} is: given observed outcomes $\mathcal U$, to construct a proxy target $Y$ that is both relevant to the target outcome $Y^*$ and reasonably predictable by predictors $\mathcal X$ through machine learning.

\begin{figure}[t]
\centering  
    \includegraphics[width=.3\textwidth]{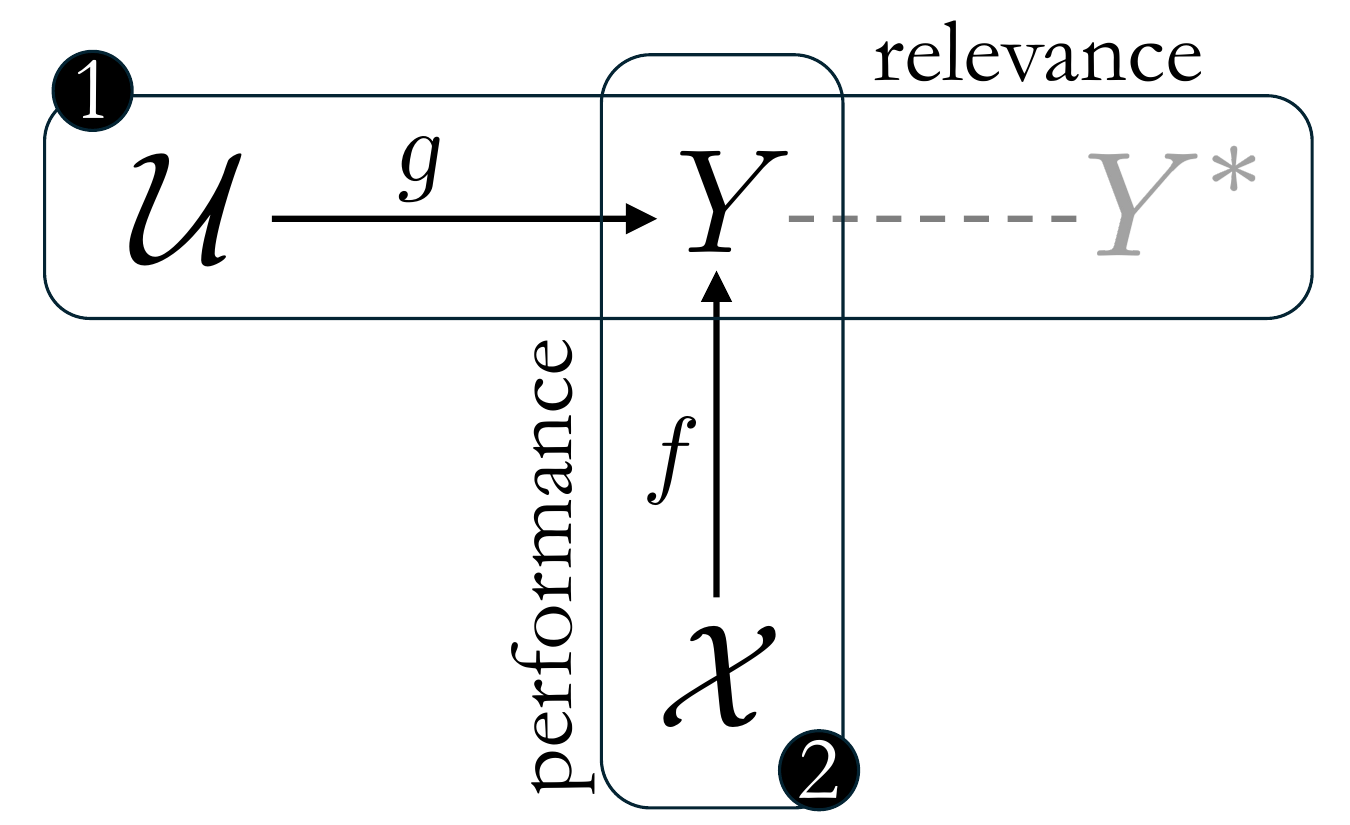}
\caption{The relationships between key concepts in the proxy target selection problem. The function $g$ uses observed outcomes $\mathcal U$ to construct the proxy target $Y$, which is a surrogate of the unobserved target outcome $Y^*$ (Box 1). The machine learning model $f$ uses predictors $\mathcal X$ to predict the proxy target $Y$ (Box 2). The problem is to construct $Y$ that is both relevant to $Y^*$ and can be accurately predicted using $\mathcal X$.}
\label{fig:formulation}
\end{figure}

The relationships between concepts in the problem are illustrated in Figure~\ref{fig:formulation}. The function $g$ in Box 1 (the horizontal rectangle) plays a central role as it produces the proxy target $Y$. $g$ cannot be trained or evaluated using data because $Y^*$ is unobserved. This is in contrast with the function $f$ in Box 2 (the vertical rectangle), which can be trained and evaluated using data because $Y$ is observed in data. The evaluation of $g$ (and therefore the proxy target $Y$) has two aspects: \emph{relevance} and \emph{performance}. The relevance aspect (whether the proxy target faithfully represents the target outcome) has to be judged based on domain knowledge in the application context, which is better suited as a human task. The performance aspect (whether the proxy target can be accurately predicted using a set of predictors) can be evaluated using standard supervised machine learning training and evaluation procedures, which is better suited as a machine task (e.g., through AutoML).

%% file: tex/experiment.tex
\section{Experimental Evaluation}

The problem formulation in Section~\ref{sec:formulation} naturally derives two solution strategies: \textsc{Relevance First} and \textsc{Performance First}. In \textsc{Relevance First}, the human user leads the process by first choosing proxy targets that achieve high relevance (Box 1 in Figure \ref{fig:formulation}), and then check their performance (Box 2 in Figure \ref{fig:formulation}). In \textsc{Performance First}, the machine leads the process by first choosing system-generated proxy targets that achieve high performance (Box 2), and then ask the human to select those that are relevant (Box 1). In this section, we evaluate the effects of these two human-machine teaming strategies (interface conditions) in a controlled within-subject user study.

\subsection{Research Questions and Hypotheses}
The study was designed to answer the following research questions (RQs) and hypotheses (Hs).

\textbf{RQ1 (Objective outcomes):} How do different human-machine teaming strategies  influence the relevance and performance of the final proxy targets selected by a user?
\begin{itemize}
    \item \textbf{H1.a (Relevance):} We hypothesize that participants would achieve similar relevance in \textsc{Performance First} and \textsc{Relevance First}. Though the \textsc{Performance First} condition inevitably introduces irrelevant proxies, we assume that users are driven to define relevant proxies and can reject irrelevant proxies.
    \item \textbf{H1.b (Performance):} We hypothesize that participants would achieve higher performance in \textsc{Performance First} condition than in \textsc{Relevance First} condition. The \textsc{Performance First} condition makes performance information accessible, which makes it easier for users to identify candidates that are both relevant and have high predictive performance.
\end{itemize}

\textbf{RQ2 (Subjective experience):} How do different human-machine teaming strategies influence a user's satisfaction and   decision-making experience in the proxy target selection task?
\begin{itemize}
    \item \textbf{H2.a (Satisfaction):} We hypothesize that participants would be more satisfied with their final proxy in \textsc{Performance First} than in \textsc{Relevance First}. The \textsc{Performance First} condition reduces the effort of experimenting with each candidates. Therefore, it is easier for the participants to identify satisfying proxies.
    \item \textbf{H2.b (Decision-making):} We hypothesize that participants would find it easier to decide which proxy to use in \textsc{Performance First} than in \textsc{Relevance First}. The \textsc{Performance First} condition makes performance information easily accessible, allowing users to compare candidates regarding both relevance and performance.
\end{itemize}

\subsection{Application Scenario and Proxy Target Selection Syntax}
\label{sec:prob_form_for_binary}
\textbf{Overall rationale}: To conduct a quantitative study in a lab setting, we designed a proxy target selection task that could be completed in a one-hour study session. Performing this task requires participants to have sufficient domain knowledge in the application area and be able to interpret the model performance in order to select the most useful model. Since domain experts are difficult to recruit, we selected application areas that college students were familiar with and restricted participants to those with experience in machine learning. While proxy target selection is often a collaborative task that involves domain experts and data scientists in practice, we adopted a simplified setting in which a single participant interacts with the system. In our study setting, participants are treated as playing the role of both data scientist and domain expert.  

\textbf{Task-specific dataset}: We employed a survey dataset examining the impact of COVID-19 on college students~\cite{aristovnik_impacts_2021}. We selected 52 outcome variables from the survey that can be used as outcome variables $\mathcal U$ and the remaining 109 variables as the model's predictors $\mathcal X$. We considered two target outcomes, corresponding to two tasks in a within-subject design involving two interface conditions (Section~\ref{sec:system_design}). The first target outcome was a binary variable $Y^{*}_{1}$ = ``whether a student's on academic performance is negatively impacted by COVID-19.'' The second target outcome was a binary variable $Y^{*}_{2}$ = ``whether a student's social and emotional life is negatively impacted by COVID-19.'' 

\textbf{Proxy target selection syntax}: The function $g$ mapping the observed outcomes to a proxy target follows a scoring syntax commonly used in decision-making contexts~\cite{singer_third_2016}. In such a syntax, multiple criteria are considered, each representing a specific aspect. 
Formally, the proxy variable $Y$ involves a chosen subset of observed outcomes $\mathcal V \subseteq \mathcal U$, $\mathcal V = \{V_1, \cdots, V_k\}$: 
\begin{align}
    Y = g(\mathcal U) = \mathbf 1\left\{ 
  \sum_{i=1}^k \mathbf 1 \left\{ V_i \ge t_i \right\} \ge t \right \} \ ,
  \label{eq:proxy_syntax}
\end{align}
where the indicator function $\mathbf 1\{z\} = 1$ if $z$ is true and 0 otherwise, $t_1, \cdots, t_k, t$ are cutoff thresholds. The proxy target is 1 if at least $t$ observed outcomes exceed their corresponding thresholds. 



To illustrate, imagine that a user tries to build a machine learning model to predict whether a student was negatively impacted by COVID-19. The following shows an example proxy target (Q20c and Q26b are identifiers of outcome variables in the dataset):

\begin{itemize}
    \item [{$V_1$} =]  Q20c (\textit{``My performance as a student has worsened since on-site classes were canceled.''}) \\ $V_1 > t_1 = 3$ means a student answered ``agree'' or ``strongly agree'' on this question. \\
     
    \item [{$V_2$} =] Q26b (\textit{``How often do you have worries about personal mental health?''}) 
    \\ $V_2 > t_2 = 4$ means a student answered ``all of the time'' on this question. \\
     
    \item [$Y$ =] $ \mathbf 1\left \{\mathbf 1 \left\{ V_1 > 3\right\} + \mathbf 1 \left\{ V_2 > 4\right\} \right \} \ge 1$. It means $Y = 1$ if either $V_1 > 3$, or $V_2 > 4$, or both.
\end{itemize}

In this case, two outcome variables are used to construct the proxy target, with one asking about academic performance and the other asking about mental health. Students are labeled as ``impacted'' ($Y=1$) if they meet at least one of the two criteria, and ``not impacted'' ($Y=0$) otherwise. 





In general, the proxy target selection syntax $g$ varies in different applications. We chose the above scoring syntax because 
(1) it is simple enough to be interpretable for our participants, and (2) it is also flexible enough to capture complex and diverse factors when defining a proxy, enabling one to consider any subset of observed outcomes and their combinations.

\subsection{Interface Conditions and System Design}
\label{sec:system_design}
\begin{figure}[t]
\centering  
    \includegraphics[width=1\textwidth]{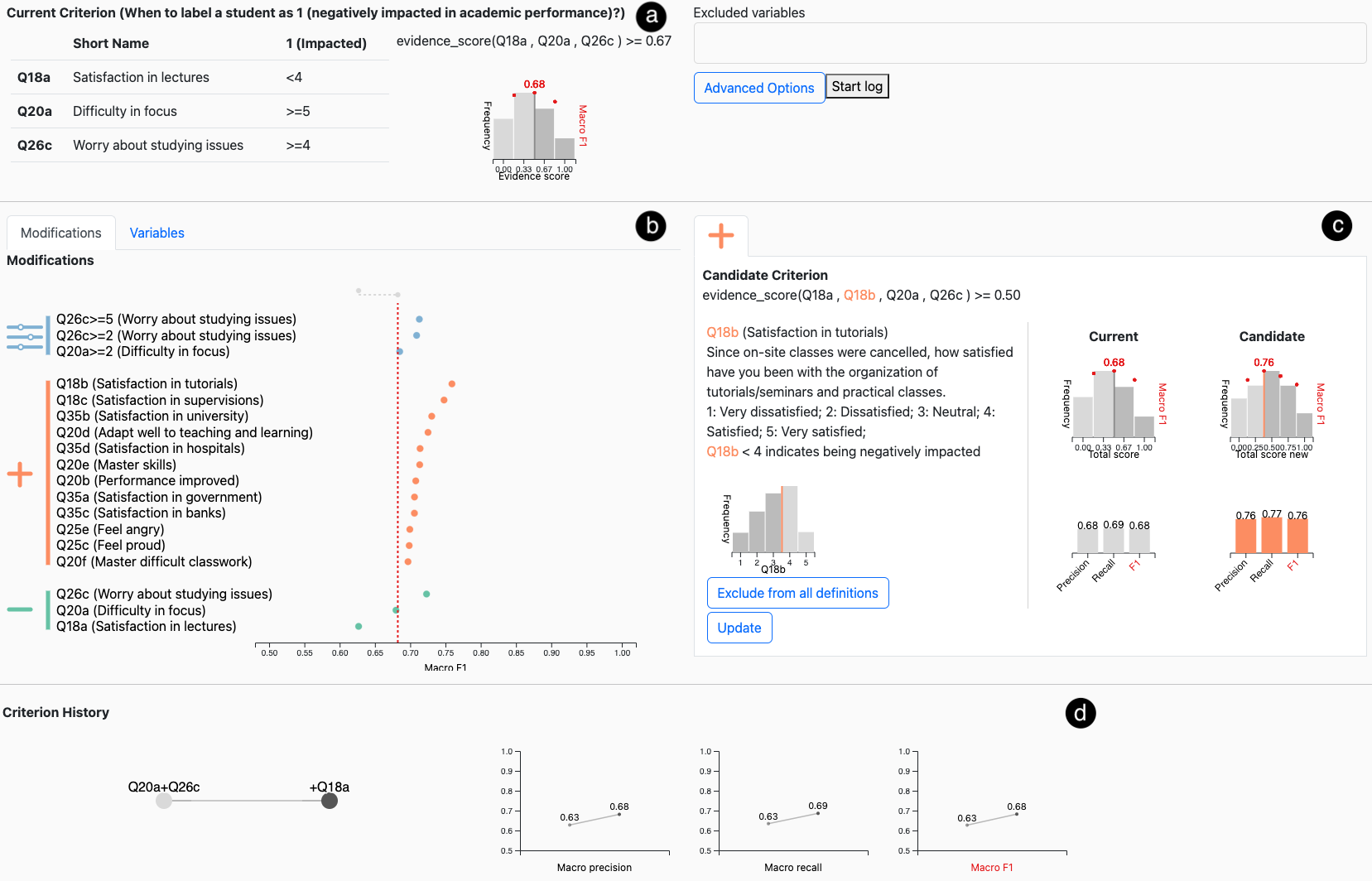}
\caption{Overview of system interface, including (a) \textit{Proxy Detail View} containing details of the current proxy target, (b) \textit{Candidate Presentation} which can take one of the two interface conditions shown in Figure~\ref{fig:system_condition_comp}, (c) \textit{Candidate Detail View} presenting the details of a selected candidate from view (b) and its comparison with the current proxy, and (d) \textit{Proxy History View} showing the iterations on proxies. As the user clicks ``Update'' in view (c), the proxy's details in view (a) will be updated and new proxy candidates will be generated and presented in view (b).}
\Description{An overview of the system. There are four views in the graph. The problem detail view is at the top of the image, with detailed information on the current proxy. Below is the candidate overview, which is a scatter plot. The x-axis is the resulting model performance of the proxy, and the y-axis is the candidate's difference from the current proxy. Each dot in the scatter plot represents a proxy candidate. To the right of the candidate overview is the candidate detail view. It shows the detailed information of the selected candidate, including the distribution and performance change from the current proxy. The history view is at the bottom of the interface. It includes a tree diagram showing the version history of proxy and line charts showing the performance changes.}
\label{fig:overview}
\end{figure}

In this section, we describe the interface customized for the \textsc{Performance First} and \textsc{Relevance First} conditions. Figure~\ref{fig:overview} shows an overview of the interface.
We assume that users start with an initial proxy target (Figure \ref{fig:overview} (a)). 
Candidates are then generated differently depending on the condition (Figure \ref{fig:overview} (b)). After a new a new proxy candidate is selected, users can view its detail and choose to update and adopt it (Figure \ref{fig:overview} (c)). Subsequently, the updated proxy target is recorded and can be backtracked (Figure \ref{fig:overview} (d)). 

\begin{figure}[t]
\centering  
\begin{subfigure}{.49\textwidth}
    \centering
    \includegraphics[width=1\textwidth]{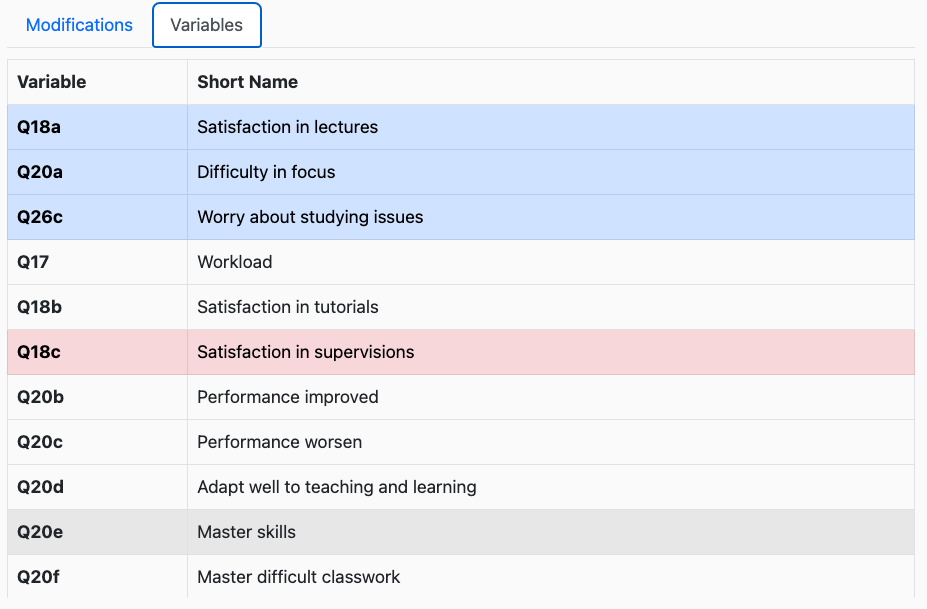}
    \caption{\textsc{Relevance First} condition}\label{fig:sys_b}
\end{subfigure}
    \hfill
\begin{subfigure}{.49\textwidth}
    \centering
    \includegraphics[width=1\textwidth]{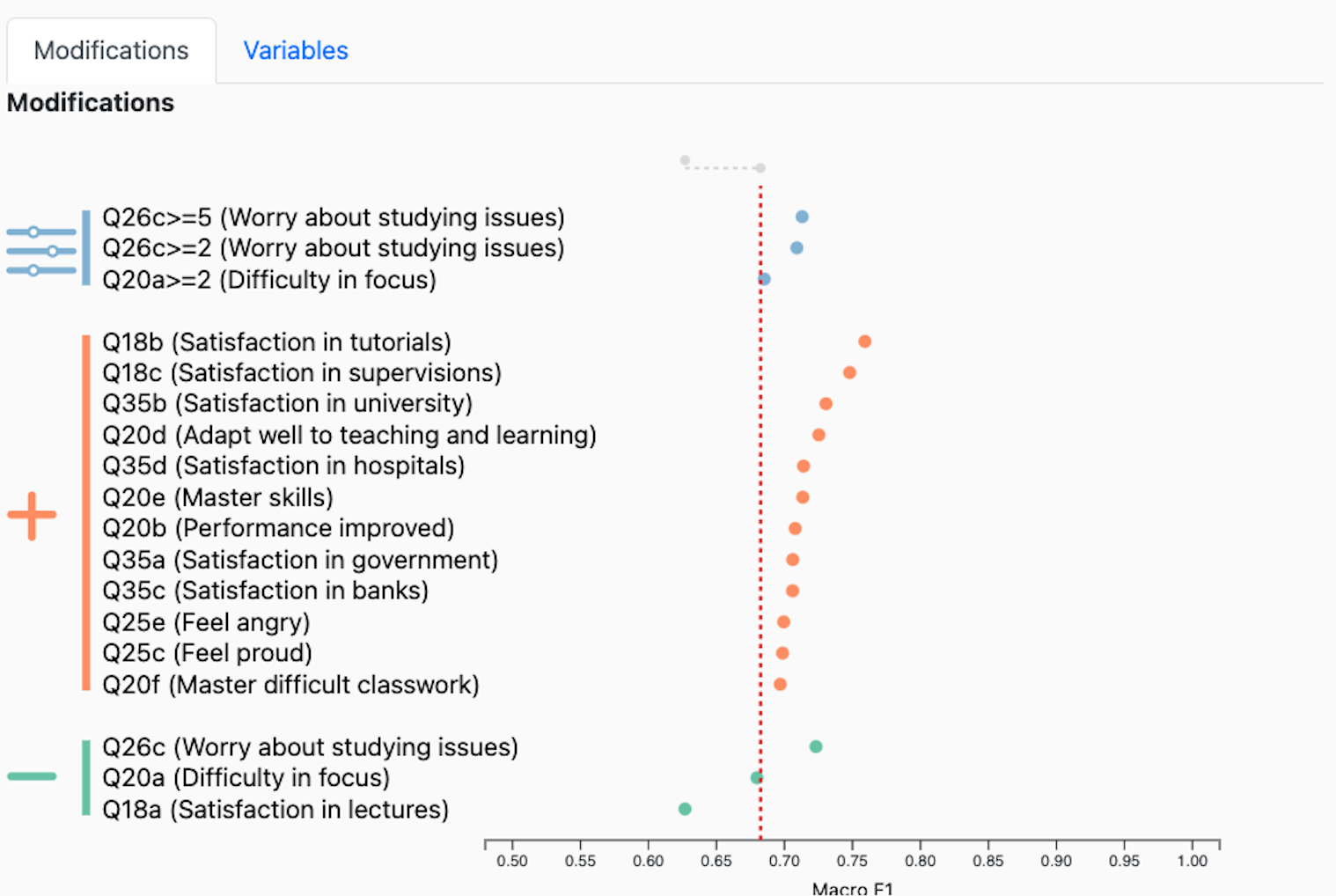}
    \caption{\textsc{Performance First} condition}\label{fig:sys_a}
\end{subfigure}

\caption{Proxy candidates presentation in the \textsc{Relevance First} condition and the \textsc{Performance First} condition. (a) \textsc{Relevance First}: all observed outcomes and those associated with the current proxy are presented in a pre-defined order based on the variables' labels. (b) \textsc{Performance First}: candidate proxies are ranked based on the resulting model performance.}
\Description{The presentation of the Relevance First and Performance First conditions. In the Relevance First condition, observed outcome variables are presented in a table. In the Performance First condition, candidate proxies are ranked by their performance. Each candidate proxy is represented as a dot in the visualization.}
\label{fig:system_condition_comp}
\end{figure}

The two interface conditions differ only in how proxy candidates are generated. 
In the \textsc{Relevance First} condition (Figure \ref{fig:system_condition_comp} (a)), the human takes the initiative to generate the next proxy candidate by interpreting the semantic meaning or relevance of observed outcomes and selecting a subset of them from a list.
In the \textsc{Performance First} condition (Figure \ref{fig:system_condition_comp} (b), which is also shown in Figure \ref{fig:overview} (b)), the machine takes the initiative to auto-generate proxy candidates and rank them by performance.

\subsubsection{Proxy Detail View}
The current proxy target variable is shown in the \textit{Proxy Detail View} (Figure \ref{fig:overview} (a)), with the included observed outcomes (criteria) and the performance of the prediction model trained for this proxy. The bar chart displays the distribution of total scores within the dataset. We normalize the total score to a range between 0 and 1 for an easier understanding of the threshold. The cutoff threshold (0.67 in Figure \ref{fig:overview} (a)), the associated model performance (F1  = 0.68), and the Class $\mathbf{1}$ population distribution (bars to the right of the cutoff in the histogram) are shown. 

By default, the system selects the cutoff threshold that maximizes the evaluation metric selected by the user. Macro-averaged F1 score is the default evaluation metric for  predictive performance.

\subsubsection{Candidate Presentation}

In the \textsc{Relevance First} condition, the system presents all observed outcome variables in a predefined order based on variables' labels (Figure~\ref{fig:system_condition_comp} (a)). All variables included in the current proxy are colored in blue, while all variables the user chooses to exclude are colored in red. When the user clicks on a new variable, the new proxy resulted from adding the variable is shown in the \textit{Candidate Detail View} (Figure \ref{fig:overview} (c)). For instance, in Figure~\ref{fig:system_condition_comp} (a), when the user clicks the variable Q20e, the proxy candidate that adds this variable to the proxy is shown. In this condition, the inspection of  proxy candidates is driven by the user's understanding of the observed outcome variables' relevance to the target outcome. The resulting performance is only shown in the \textit{Candidate Detail View} after a set of observed outcome variables are selected.

In the \textsc{Performance First} condition, the system auto-generates a set of proxy candidates (details below), ranks them based on the predictive model performance, and presents the top-performing candidates. For an effective overview of the proxy candidates, the system displays the candidates in a 2D scatter plot, where the x-axis displays the model performance measure (F1 score as shown in Figure \ref{fig:system_condition_comp} (a)) and the y-axis displays the details of the applied modification, which allows the user to make relevance judgments based on each variable's meaning. 

Candidates are grouped by modification types (details below). Candidates with the same modification type are ranked by their model performance. The system presents at least three candidates in each modification group and no more than 21 candidates across modification groups. These numbers are selected to ensure users have an ample and diverse list of choices for selection while avoiding overwhelming information. The red vertical dash line indicates the performance of the current proxy.

Any proxy candidate (a dot in the 2D scatter plot) can be selected with a click. The selected candidate is shown with more details in the \textit{Candidate Detail View} (Section \ref{sec:cand_detail_view}). Users may exclude candidates related to certain variables from the list to focus on interesting candidates. 

\textbf{Candidate generation in the \textsc{Performance First} condition}: The space of possible proxies grows exponentially with the number of observed outcome variables, which can be overwhelming for the user to explore. In addition, it is computationally infeasible to exhaustively explore on all potential proxies at once. To reduce the number of new proxies explored in each iteration, the system generates candidates that are similar to the current proxies. This allows users to make gradual, iterative refinements to their proxies. Specifically, the system enumerates all candidates that are \emph{within the one-step edit distance} from the current proxy.

Formally, given a proxy target $Y$ as shown in Equation \eqref{eq:proxy_syntax} that involves a subset of outcome variables $\mathcal V \subseteq \mathcal U$,  we considered three types of one-step modifications: 
\begin{itemize}
    \item [\textbf{(1)}] \textbf{Threshold Change:} Modify the threshold of an existing criterion in $\mathcal V$, e.g., $V_1>t_1 \rightarrow V_1>t_1'$.
    \item [\textbf{(2)}] \textbf{Addition:} Add a new outcome variable $V_{k+1} \in \mathcal U \backslash \mathcal V$ to $Y$ with a new term $\mathbf 1\left\{V_{k+1}>t_{k+1}\right\}$. Here, $t_{k+1}$ is set to the default threshold of $V_{k+1}$.
    \item [\textbf{(3)}] \textbf{Deletion:} Delete an outcome variable from $Y$.
\end{itemize}

Given a proxy target $Y$, the system trains and evaluates the corresponding prediction model $f$ on a training-test data split. In our case, we train and evaluate a logistic regression model due to its efficiency and decent predictive performance. In principle, the system can be extended to other ML models and AutoML techniques that automate the model family and hyperparameter selection processes.


\subsubsection{Candidate Detail View}
\label{sec:cand_detail_view}
Detailed information about the selected candidate is shown in the \textit{Candidate Detail View} (Figure \ref{fig:overview} (c)). Details of the modified variable are displayed on the left for user insight into the candidate's validity. The comparison between the current and candidate formulation in terms of distribution and performance is shown on the right-hand side for performance evaluation. 



\subsubsection{Proxy History View}
\textit{Proxy History View} provides a visual overview of different versions of proxy targets in the format of a history tree (Figure \ref{fig:overview} (d)). Each node in the history tree represents one version of the proxy target, and the text above indicates the applied modification. A user may return to a previous version by clicking on a node. The change in model performance is shown as a line chart on the right-hand side. 


\subsection{Study Design}
\subsubsection{Overview}
\label{sec:study_overview}

As illustrated in Table \ref{tab:study_design}, we adopt a within-subject design ($N=20$) where each participant conducted two study sessions, each with a different interface condition (\textsc{Performance First} and \textsc{Relevance First}) and a different target outcome (``whether a student's academic performance was negatively impacted by COVID-19'' and ``whether a student's social and emotional life was negatively impacted by COVID-19''). We counter-balanced the presentation order of the interface conditions and the target outcomes across subjects, as illustrated in Table~\ref{tab:study_design}. To control for differences in starting points, we pre-specified the initial proxy for each application scenario such that they are plausibly relevant with respect to the target outcome and have a relatively low predictive performance.


\begin{table}[H]
\small
\begin{tabular}{c|cc|cc}
\toprule
        & \multicolumn{2}{c|}{Sub-session 1}             & \multicolumn{2}{c}{Sub-session 2}             \\ \midrule
        & \multicolumn{1}{c|}{Target Outcome}      & Interface Condition   & \multicolumn{1}{c|}{Target Outcome}      & Interface Condition   \\ \midrule
Group 1 & \multicolumn{1}{c|}{Academic Performance}   & \textsc{Performance First}    & \multicolumn{1}{c|}{Social and Emotional Life} & \textsc{Relevance First} \\ 
Group 2 & \multicolumn{1}{c|}{Social and Emotional Life} & \textsc{Relevance First} & \multicolumn{1}{c|}{Academic Performance}   & \textsc{Performance First}    \\ 
Group 3 & \multicolumn{1}{c|}{Social and Emotional Life} & \textsc{Performance First}    & \multicolumn{1}{c|}{Academic Performance}   & \textsc{Relevance First} \\ 
Group 4 & \multicolumn{1}{c|}{Academic Performance}   & \textsc{Relevance First} & \multicolumn{1}{c|}{Social and Emotional Life} & \textsc{Performance First}    \\ \bottomrule
\end{tabular}
\caption{Experimental design. Participants were split into four groups. We counter-balanced the order of using the two systems and the scenarios used for testing the two systems.}
\label{tab:study_design}
\end{table}


\subsubsection{Proxy Target Selection Tasks}
\label{sec:problem_formulation_tasks}
Participants were asked to imagine that they needed to develop an ML model that predicts whether COVID-19 had negative impacts on students' academic performance or social and emotional life. Given the the application scenario and dataset described in Section \ref{sec:prob_form_for_binary}, participants were tasked to refine the initial proxy to fulfill two objectives: 1) the proxy should be relevant to the modeling goal, encoded using a representative and comprehensive set of observed outcome variables, and 2) the final proxy should achieve better predictive performance than the initial proxy if possible. 

The selection of the dataset and the application scenarios was guided by two primary goals: 1) The dataset should encompass a diverse array of observed outcome variables so that a wide variety of relevant proxies can be constructed, and 2) the target outcomes (constructs of interest) should be familiar to the participants. We used the same dataset for the two tasks in the study to minimize the difference between the two tasks. We selected the two application scenarios with minimal overlap in  relevant observed outcome variables to reduce the learning effect in a within-subject design.

\subsubsection{Participants}
We recruited 20 participants (11 males, 9 females) via campus-wide mailing lists. All participants had either taken a course or completed an online tutorial on machine learning. Eleven participants' field of study was information science, six participants' field of study was computer science, and three participants' field of study was statistics. Participants were randomly assigned to different groups shown in Table~\ref{tab:study_design}. 

\subsubsection{Procedure}
\label{sec:procedure}
In-person user study sessions lasted roughly one hour. After providing informed consent, we provided detailed information about the application context, dataset, and the proxy target selection syntax. Each study session included two sub-sessions. We began each sub-session with a hands-on tutorial on the corresponding interface condition using an example target outcome. Participants were then informed of the application scenario to work on and manually judged the relevance of a subset of observed outcome variables with respect to the target outcome. Then, participants completed the proxy target selection task and was encouraged to think aloud in the process. To ensure that participants took the task seriously, we asked them to justify their proxy after each sub-session. Participants could spend at most 15 minutes on each task. Afterwards, participants completed a post-task questionnaire (Table~\ref{tab:questionnaire}) and answered interview questions to provide feedback. Each participant was rewarded an Amazon gift card worth US\$20  for participating in the study. The study was approved by our Institutional Review Board (IRB).


\begin{table}[]
\small
\begin{tabular}{lll}
\toprule & & Question\\ \midrule
H2.a & & \begin{tabular}[c]{@{}l@{}}Reflecting on the process of constructing your final criterion, how \\ difficult it was for you to:\end{tabular} \\ \cmidrule{2-3} 
& Q1 Relevance  & Select variables relevant to the current topic \\ \cmidrule{2-3} 
& Q2 Completeness & \begin{tabular}[c]{@{}l@{}}Select a set of variables that adequately cover the aspects relevant \\ to the current topic\end{tabular}       \\ \cmidrule{2-3} 
& Q3 Performance & Come up with a proxy that reaches a satisfying performance  \\ \cmidrule{2-3} 
& Q4 Overall & Come up with an overall satisfying proxy \\ \midrule
H2.b & & \begin{tabular}[c]{@{}l@{}}During the task, to what extent do you think the interface helped \\ you in the following aspects:\end{tabular}  \\ \cmidrule{2-3} 
 & \begin{tabular}[c]{@{}l@{}}Q5 Performance \\       Difference\end{tabular} & See the performance difference between modifications \\ \cmidrule{2-3} 
& \begin{tabular}[c]{@{}l@{}}Q6 Semantic \\       Difference\end{tabular}    & See the semantic difference between modifications \\ \cmidrule{2-3} 
& Q7 Decision & Decide which modification to select at each step \\ \cmidrule{2-3}
& \begin{tabular}[c]{@{}l@{}}Q8 Performance \\       v.s. Semantic\end{tabular} & Rate the relative importance between performance and semantic in your decision \\
\bottomrule
\end{tabular}
\caption{Post-Task Questionnaire Questions}
\label{tab:questionnaire}
\end{table}

%% file: tex/results.tex
\section{Results}

\subsection{RQ1: Objective Outcomes}

\label{sec:questionnaire_h1}
\begin{figure}[t]
\centering  
\begin{subfigure}{.49\textwidth}
    \centering
    \includegraphics[width=1\textwidth]{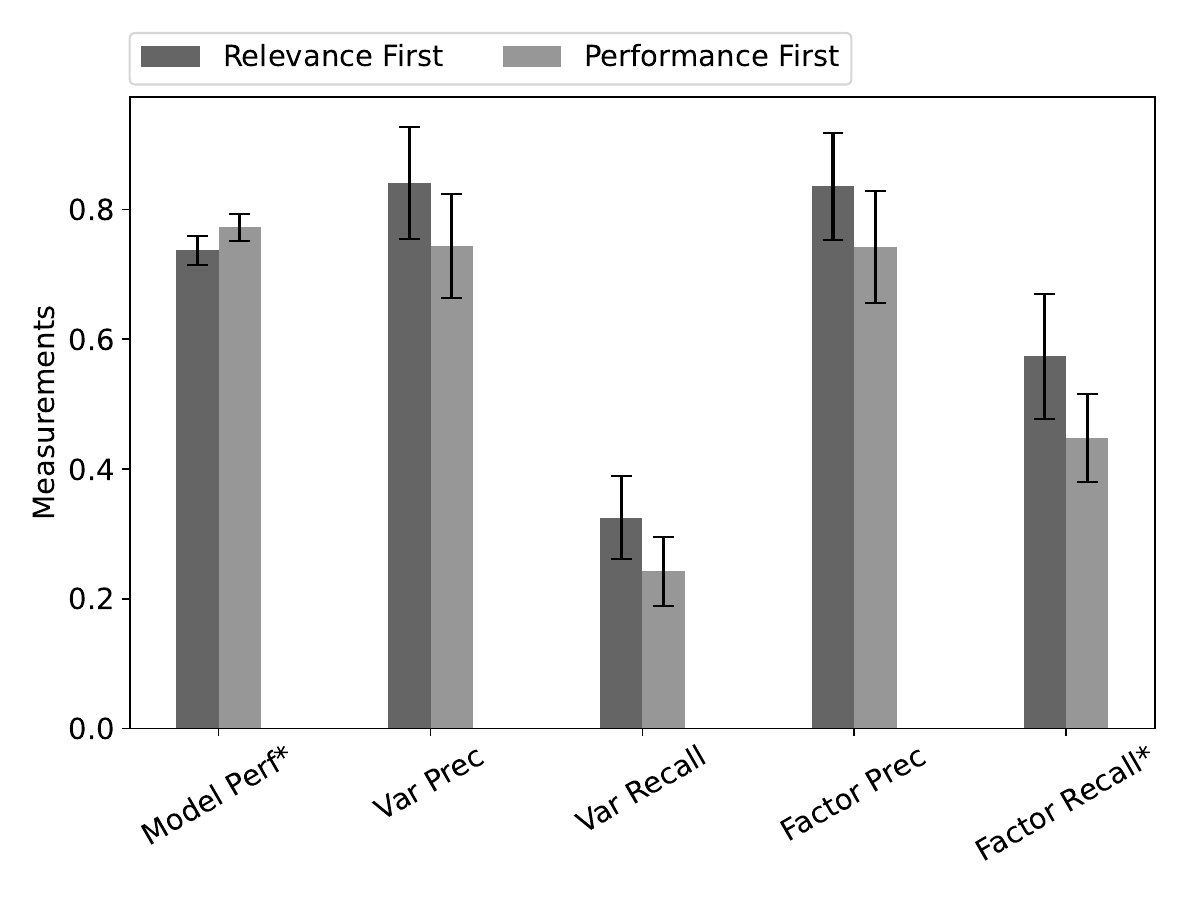}
    \caption{Relevance and performance of final proxy targets}\label{fig:res_a}
    \Description{Bar chart showing the objective quality measurements of proxy targets generated using the \textsc{Relevance First} and \textsc{Performance First} conditions. The graph shows that the \textit{Performance First} condition produced proxy targets with significantly higher model performance than the \textsc{Relevance First} condition, while the \textsc{Relevance First} condition produced proxy targets with significantly higher factor recall than the \textit{Performance First} condition.}
\end{subfigure}
    \hfill
\begin{subfigure}{.49\textwidth}
    \centering
    \includegraphics[width=1\textwidth]
    {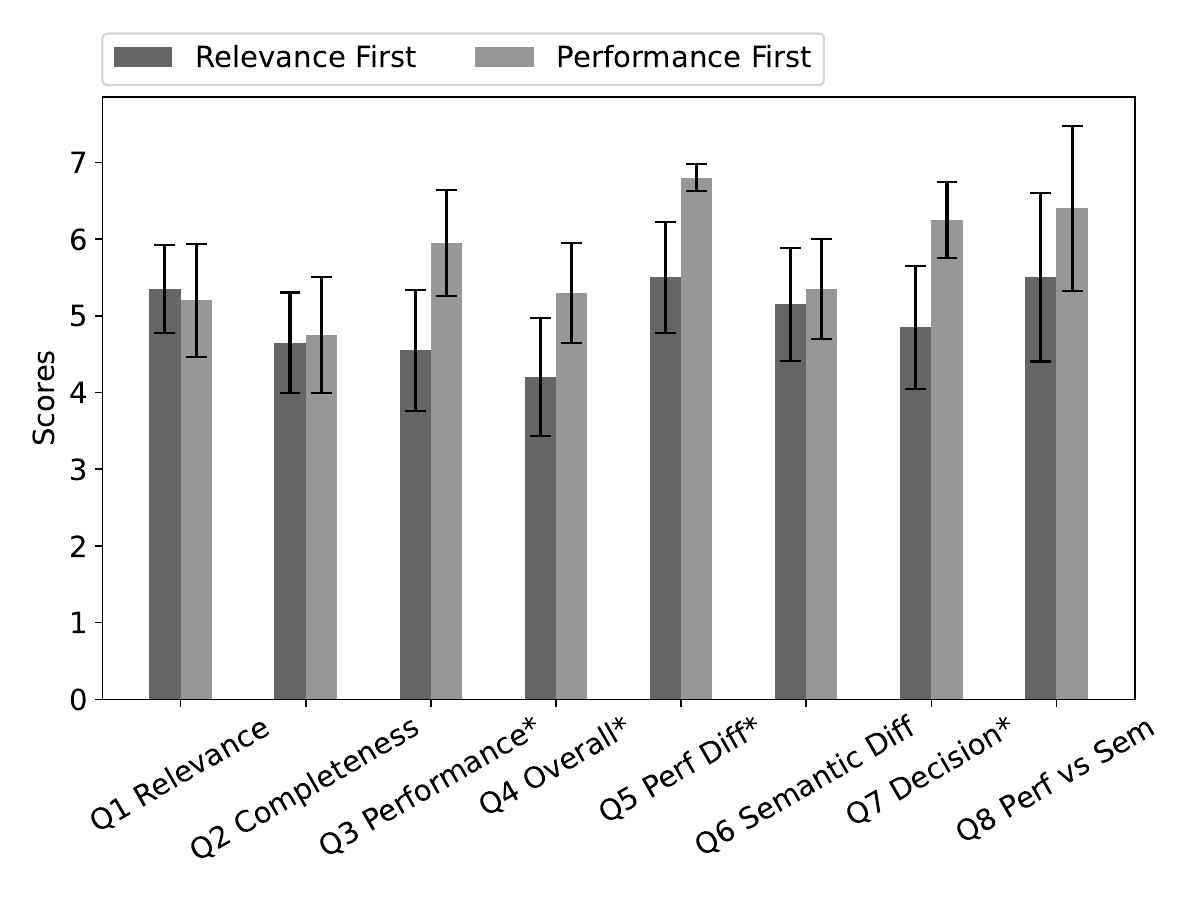}
    \caption{Questionnaire results}\label{fig:res_b}
    \Description{Bar chart showing the questionnaire responses on Q1-Q5.  The graph shows that participants give significantly higher ratings to the \textit{Performance First} condition on Q4 Overall and Q5 Performance Difference.}
\end{subfigure}

\caption{(a) Quality measurements of proxies generated by participants under \textsc{Performance First} and \textsc{Relevance First} conditions. (*) indicates statistically significant differences between the two conditions ($p<0.05$). There is a significant difference in the resulting model performance of the proxies generated under the two conditions (\textsc{Performance First}>\textsc{Relevance First}). There is a significant difference in factor recall, but no significant difference in variable precision, recall, and factor precision. (b) Mean ratings given by participants in the post-task questionnaire. Error bars show the standard error. There is a significant difference between \textsc{Performance First} and \textsc{Relevance First} conditions in Q3 Performance, Q4 Overall, Q5 Performance Difference, and Q7 Decision (see Table~\ref{tab:questionnaire} for question details).}
\label{fig:res1}
\end{figure}
Each participant produced two proxy targets using the scoring syntax, one for each condition. The objective quality of the proxy targets, including relevance and performance, was measured and compared between conditions. Proxy quality measurements were averaged within system conditions for comparison between conditions. Fisher’s randomization test ($\alpha = 0.05$) was used to test for significant effects due to system differences. 

We leveraged a model's predictive performance to measure the performance of a proxy. Considering the potential imbalance between Class $\mathbf{1}$ and Class $\mathbf{0}$, we employed macro-averaged F1 score as the performance metric, which is also the optimization target for participants during the study. 


Relevance is challenging to measure as the task was designed to be open-ended and there are multiple aspects of relevance, such as topicality, coverage, and redundancy. To cope with these challenges, we made use of participants' assessments of variables' relevance to target outcomes. We evaluated a proxy target's relevance by analyzing the variables involved in it.

As described in Section \ref{sec:procedure}, participants judged the relevance of a set of variables with respect to the application scenario before each task. We obtained the relevance of each observed outcome variable $r(U_i)$ by aggregating participants' relevance judgments. In addition, we conducted a factor analysis to find the underlying constructs in the outcome variables, which resulted in eleven factors $\mathcal Z = \{Z_j\}_{j=1}^{11}$. The factor Relevance $r(Z_j)$ is calculated by averaging the relevance scores of variables included in the factor. 

We then make use of $r(U_i)$ and $r(Z_j)$ to evaluate a proxy target's relevance on four metrics. Given a proxy target  represented by a subset of observed outcome variables $\mathcal V = \{V_1, \cdots, V_k\}$ and a set of factors $\mathcal Z$, a list of variables relevant to the target outcome $\mathcal V_{rel}$, and a list of factors relevant to the target outcome $\mathcal Z_{rel}$, we calculated four measurements: 
\textbf{Variable Precision} $ \frac{|\mathcal V \cap \mathcal V_{rel}|}{|\mathcal V|}$,
\textbf{Variable Recall} $ \frac{|\mathcal V \cap \mathcal V_{rel}|}{|\mathcal V_{rel}|}$,
\textbf{Factor Precision} $\frac{|\mathcal Z \cap \mathcal Z_{rel}|}{|\mathcal Z|}$, 
and \textbf{Factor Recall} $\frac{|\mathcal Z \cap \mathcal Z_{rel}|}{|\mathcal Z_{rel}|}$. These metrics evaluate the preciseness and completeness of the observed outcome variables included in a proxy. In this study, we use these metrics to approximate the relevance of the proxy. 

The evaluation results of the proxy quality, including the resulting model's performance and the four measurements for relevance, are shown in Figure \ref{fig:res_a}. 

Participants achieved significantly higher model performance when under \textsc{Performance First} ($Avg_{perf}=0.773$, $Avg_{rel}=0.737$, $p = 0.034$). This result \textbf{supports} \textbf{H1.b (Performance): the \textsc{Performance First} condition helped users produce proxy targets that have higher predictive performance}. However, participants reached an overall lower relevance score under the \textsc{Performance First} condition.
There is no significant difference in variable precision ($Avg_{perf}=0.743$, $Avg_{rel}=0.841$, $p = 0.152$), variable recall ($Avg_{perf}=0.242$, $Avg_{rel}=0.325$, $p = 0.064$), and factor precision ($Avg_{perf}=0.743$, $Avg_{rel}=0.836$, $p = 0.14$). However, participants reached significantly lower factor recall in the \textsc{Performance First} condition than in the \textsc{Relevance First} condition ($Avg_{perf}=0.448$, $Avg_{rel}=0.574$, $p = 0.029$), i.e., less concept groups were included in the proxy when using \textsc{Performance First}. We also observed lower scores in other relevance metrics, though not significant. These results thus \textbf{refute H1.a (Relevance): participants did \emph{not} achieve a similar level of relevance under the two conditions}. 

In summary, the \textsc{Performance First} condition led participants to select well-performing proxies, even though they are less relevant to the modeling goal and may not cover all aspects of the construct of interest. One potential explanation is that participants were influenced by the ranking, as candidate proxies were ranked by the resulting model's performance in the \textsc{Performance First} condition. Another possible explanation is that the bias towards well-performing proxies exists in both conditions. The two conditions expose participants to different sets of candidates, leading to different selection results.

To understand what led to the observed differences between conditions, we investigated the proxy candidates checked versus selected in each step of modification. We focus on analyzing two quality metrics, the resulting model's performance and variable precision. Within one step of modification, participants checked multiple candidate proxies by either clicking variables (\textsc{Relevance First} and \textsc{Performance First}) or looking at the list of candidates  recommended by the system (\textsc{Performance First}). They then selected a proxy as the updated proxy, which initiated the next modification step. Within one modification step, we recorded the maximum and minimum of model performance and variable precision of proxies checked by each participant. This helps us understand what candidates were checked and how those candidates influenced a participant's final choices. 

\begin{figure}[t]
\centering  
\begin{subfigure}{.49\textwidth}
    \centering
    \includegraphics[width=1\textwidth]{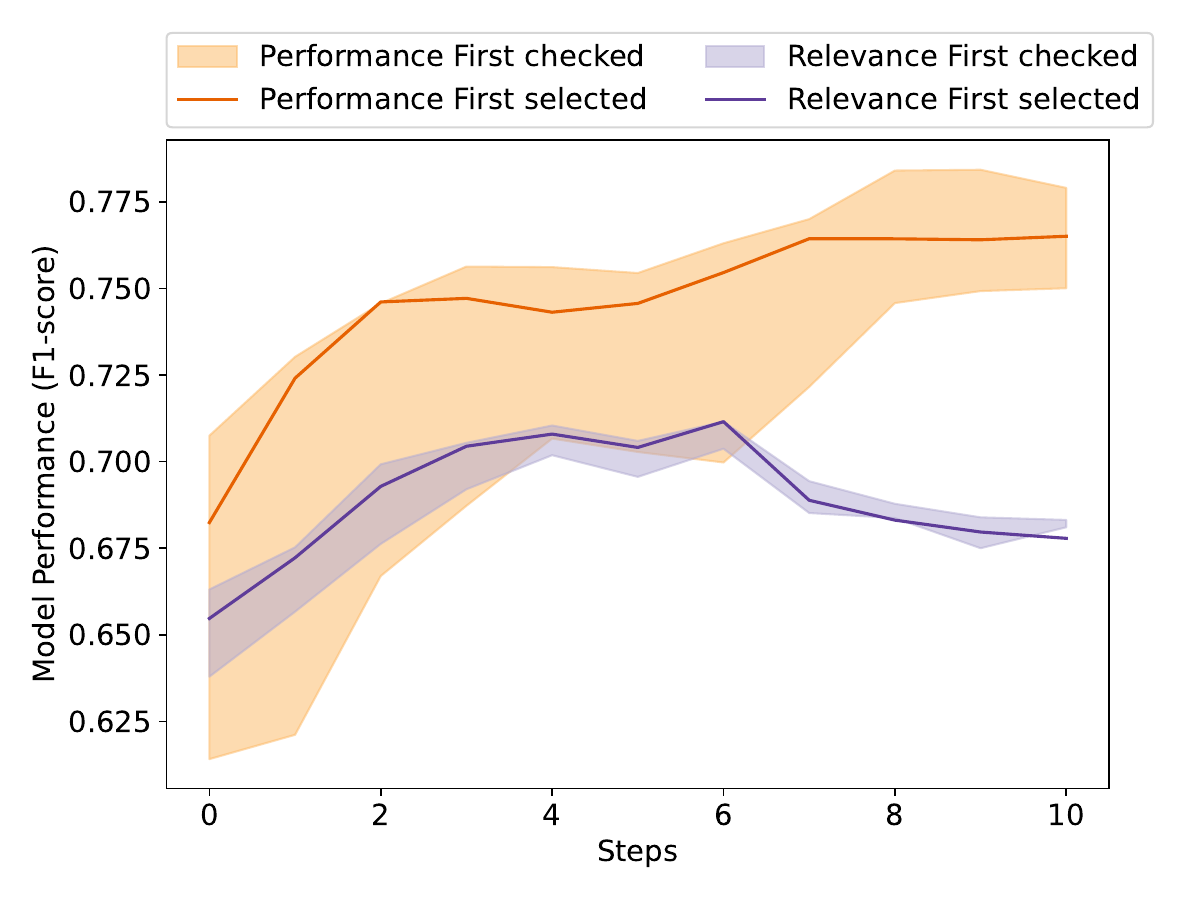}
    \caption{Change of model performance during study sessions}\label{fig:progress_perf}
\end{subfigure}
    \hfill
\begin{subfigure}{.49\textwidth}
    \centering
    \includegraphics[width=1\textwidth]
    {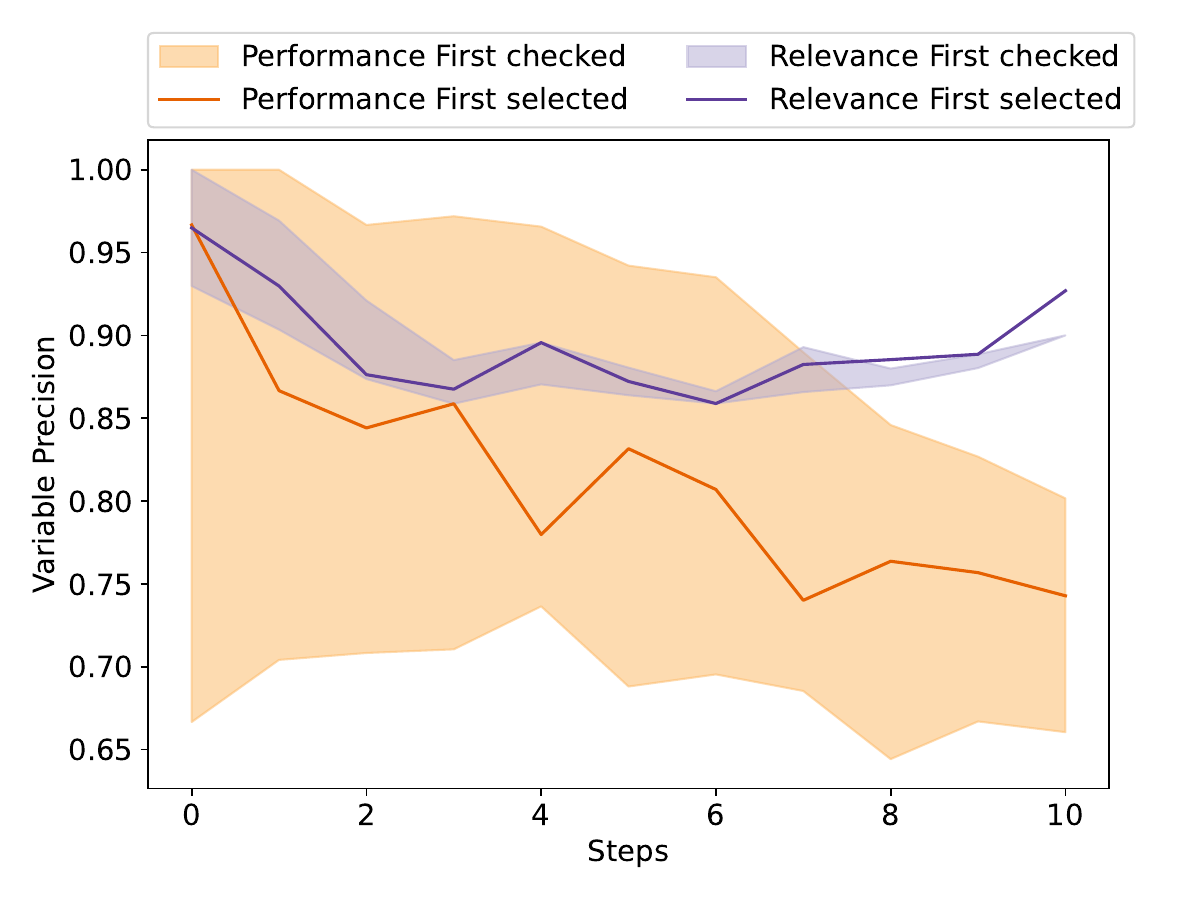}
    \caption{Change of proxy variable precision during study sessions}\label{fig:progress_rel}
    
\end{subfigure}

\caption{The change of model performance and proxy variable precision during the study session averaged across all participants. The solid lines indicate the selected proxy during each iteration. The shadow areas indicate the range of performance and variable precision of the proxies examined by participants in each iteration. Participants tend to select the proxy that results in higher performance relative to the set of candidates examined in both conditions. Participants were exposed to a wider range of candidates in \textsc{Performance First} than in \textsc{Relevance First}.}
\label{fig:progress}
\end{figure}

In Figure \ref{fig:progress}, we plotted in total 11 steps of modifications. We chose 11 steps because half of the participants have done 11 or more steps of modifications. The solid lines indicate model performance or variable precision of the chosen proxy in each step, averaged across all participants. The shaded areas indicate the range of model performance and variable precision of the proxies checked by participants in each step, averaged across all participants. Note that in the \textsc{Performance First} condition, we assume all candidates listed in the ranked list were checked by participants.
 
Several trends can be observed in Figure \ref{fig:progress}. Under \textsc{Performance First}, participants updated their proxies to increase the model performance. At the same time, their variable precision decreased. Under \textsc{Relevance First}, a similar trend was observed in the first five modification steps. Then, an increase in variable precision and a decrease in model performance were observed, hinting that participants might focus more on improving the relevance of the proxy target variables. These results suggest that \textbf{under \textsc{Performance First}, participants were mainly driven to improve the model's performance}, while \textbf{under \textsc{Relevance First}, participants attempted to both improve model performance and align the proxy with the modeling goal}.

It can also be observed that participants tend to select well-performing candidates among the candidates checked under both conditions. In the first seven steps of modification in Figure \ref{fig:progress_perf}, the solid lines are located near the upper bound of the shaded area,  indicating that the selected proxies' performance is close to the maximum of the proxies checked. This demonstrates that  \textbf{\emph{performance bias}, i.e., participants' tendency to select well-performing proxies, might exist in both conditions}. Under \textsc{Relevance First}, participants selectively examined candidates that were more relevant to the modeling goal (see Figure \ref{fig:progress_rel}). However, the \textsc{Performance First} condition presented participants with a larger set of candidates, including those with very high model performance. In this case, performance bias led participants to select candidates that resulted in higher model performance, though they were less aligned with the modeling goal.

The above analysis demonstrates the potential negative effect of the faster iterations enabled by AutoML --- exposing users to a wider range of options might enlarge the effect of bias towards well-performing proxies. This bias might be due to the fact that model performance is easier to quantify and optimize compared to evaluating whether a proxy is relevant to the modeling goal. In this study, the \textsc{Performance First} condition exposes participants to more proxies that exhibit high performance, leading to higher model performance and lower proxy relevance. The \textsc{Performance First} condition may also implicitly encourage participants to optimize model performance instead of paying attention to a proxy's alignment with the modeling goal. 
We instructed participants to maintain and improve a proxy's relevance to the modeling goal under both conditions. 
Though participants were aware of this objective, as demonstrated by the \textsc{Relevance First} condition, there is no trend for participants to work on improving the relevance of the resulting proxy in the \textsc{Performance First} condition.

\subsection{RQ2: Subjective Experience}

In the post-task questionnaire after each sub-session, participants provided their level of agreement with statements listed in Table \ref{tab:questionnaire} on a Likert scale from 1 (strongly disagree) to 7 (strongly agree) \cite{brooke_sus_2013}. Participants' questionnaire responses were averaged within system conditions for comparison between conditions. Fisher’s randomization test ($\alpha = 0.05$) was used to test for significant effects due to system differences. 

As shown in Figure \ref{fig:res_b}, participants thought it was easier to reach a satisfying performance (Q3, $Avg_{perf}=5.95$, $Avg_{rel}=4.55$, $p = 0.014$) and reach an overall satisfying proxy (Q4, $Avg_{perf}=5.3$, $Avg_{rel}=4.2$, $p = 0.017$) in the \textsc{Performance First} condition than the \textsc{Relevance First} condition. There is no significant difference in terms of reaching satisfying results in the validity aspects (Q1, Q2). Participants perceived that in the \textsc{Performance First} condition it was easier to see the performance difference (Q5, $avg_{perf}=6.8$, $avg_{rel}=5.5$, $p=0.006$), and decide which proxy to choose (Q7, $avg_{perf}=6.25$, $avg_{rel}=4.85$, $p=0.002$) comparing to the \textsc{Relevance First} condition. There is no significant difference in understanding the semantic meaning of the candidates under the two conditions.

Indeed, \textbf{the \textsc{Performance First} condition led to more satisfying proxies and easier decision-making}, thus supporting \textbf{H2.a (Satisfaction)} and \textbf{H2.b (Decision-making)}. However, our findings from RQ1 imply that this condition may have also introduced a cognitive shortcut: by emphasizing predictive performance, participants may become overly focused on this single metric. As a result, they  felt more satisfied and reached decisions more easily, even when the chosen proxies were less aligned with the modeling goal.



%% file: tex/discussion.tex
\section{Discussion and Implications}

\subsection{Benefits and Risks of Introducing Automation into Proxy Selection}

In this work, we showed that the task of proxy target selection in machine learning applications presents a new opportunity for human-machine teaming. We explored the effects of different interface conditions to facilitate proxy target exploration. Through a controlled user study, we observed the benefits and risks of the \textsc{Performance First} and \textsc{Relevance First} strategies. 
The first benefit of human-machine teaming is the speed. In both conditions, participants were able to finish on average 10 iterations within a 15-minute study session. In current practice, 10 iterations may take a team months to complete. The \textsc{Performance First} strategy made it easier for participants to achieve high model performance than the \textsc{Relevance First} condition. Participants were exposed to a larger set of systematically generated proxy options than what they might  manually try in \textsc{Relevance First}. This larger set of options might contain proxies that were both relevant and well-performing, waiting to be discovered. 

Despite these opportunities, there is also evidence showing risks of selecting not-so-relevant proxies in the \textsc{Performance First} condition. Based on the questionnaire results, there is no significant shift in participants' perceived importance of performance and relevance. However, the proxies selected under the \textsc{Performance First} condition show a lower level of relevance. As remarked by some of our participants, one can experience a ``performance bias,'' i.e., the tendency to pursue well-performing as opposed to relevant proxies, as performance metrics are directly measurable and visible, as opposed to the fuzzy and abstract notion of relevance. While semantic relevance requires subjective judgments, model performance is quantified, making it easily available for comparison. The difference in how the objectives are quantified implicitly adds more weight to performance in people's decision-making. Notably, further analysis showed that ``performance bias'' probably existed in both \textsc{Performance First} and \textsc{Relevance First}. Since participants were exposed to more options under the \textsc{Performance First} condition, the effect of performance bias could be enlarged.


The phenomena of ``performance bias'' observed in our study may be interpreted as a form of over-reliance on machine-generated outputs~\cite{buccinca2021trust, chiang2021you}. Model performance metrics are often presented as objective measurements provided by the system. In contrast, the relevance of a proxy relies more on users' subjective judgment. The difference in the perceived objectivity can encourage users to rely on performance metrics when making decisions. Our finding suggests that in multi-criteria decision-making scenarios like proxy target selection, quantifying a single objective and recommending candidates based on it might influence how users weigh different objectives in their decision-making.

\subsection{Future Directions for Proxy Selection Support Systems}

The above findings suggest two avenues for future research. First, how can we quantify the relevance of candidate proxies and incorporate this into users' search processes? This can potentially mitigate the bias towards performance by making the evaluation and comparison of relevance equally quantified. Furthermore, quantifying relevance would enable the use of multi-criteria optimization methods to balance performance and relevance in the search process. However, quantifying whether a proxy aligns with the construct of interest is challenging, as the construct of interest is often ambiguous and difficult to fully define. Besides semantic similarity, tools from measurement and modern validity theory such as convergent validity, discriminant validity, and predictive validity, could be applied to evaluate relevance~\cite{coston_validity_2023, Guerdan_groundless_2023}.

Second, instead of quantifying the relevance objective, can we present the performance information differently to encourage a holistic understanding of model performance regarding the selected proxy? Techniques such as instance-based and data slice-based model evaluation can effectively communicate what the model actually learned and how significant a wrong prediction could be~\cite{polyzotis2019slice, zhang2022sliceteller}. Future research could investigate ways to present these qualitative evaluation results and assess their impact on proxy selection.

\subsection{Supporting Collaborative Proxy Selection}

In this study, we investigated the effect of human-machine collaboration on proxy selection. In our study design, the system mainly served the role of a data scientist, while study participants mainly serve the role of a domain expert, who also have the ability to interpret model performance metrics. This provides an easy lab setting for us to study the effect of different conditions. However, we also see the potential for applying our study conditions to a collaborative proxy selection process between data scientists and domain experts. Though Sivaraman et al.  \cite{sivaraman2025tempo} have studied the effect of Tempo, a system that enabled faster problem formulation iterations and easier performance interpretations, their study was qualitative. One interesting future direction is to quantitatively understand the effect of such systems during problem formulation on data science teams. 

In a collaborative setting, the system's effects may differ: data scientists can provide transparency around performance metrics, such as conducting detailed analysis of data slices and model behavior. Domain experts can provide domain knowledge independently without being influenced by model performance information. The system may prompt explicit negotiation between the competing objectives of the two roles, which may reduce performance bias. 


\section{Limitations}

Our user study has three main limitations. First, application scenarios have a large impact on the realism of the task and the  effects of interface conditions. To cope with this, we counter-balanced the scenarios participants worked on using either system. Second, proxy selection requires complex decision-making that involves deliberation of different factors. It also requires the participants to make sense of the application context and the underlying data. The limited in-lab study time (about one hour) may discourage the participants from thinking deeper. In future work, it would be valuable to study the impact of human-machine teaming by observing people using the system in the long run on real-world problems. Third, our evaluation of proxies' relevance is based on participants' relevance judgments. This method reduces the relevance to using the right set of variables, ignoring the potential interactions among different variables in the set.  In future work, other ways of relevance evaluation, such as convergent validity~\cite{convergentvalidity}, can be considered in addition to the current method.


%% file: tex/conclusion.tex
\section{Conclusion}
In this paper, we studied the scenario where machine learning practitioners need to define an appropriate proxy target variable to operationalize a construct of interest. We reported findings from a comparative study where two human-machine teaming strategies, \textsc{Performance First} and \textsc{Relevance First}, were used for proxy target selection. Our main finding is that the \textsc{Performance First} strategy allows users to achieve higher model performance, but can also bias users towards less relevant proxies. Under \textsc{Performance First}, where users are guided through the search process by performance, participants reported easier decision-making and higher satisfaction. We hope that by showing the risks and opportunities in tackling this important problem, our work will inspire the HCI community to further develop human-centered approaches to support systematic exploration and holistic evaluation of proxy targets, such that models in high-stakes machine learning applications will ``predict the right proxy'' and ``predict the proxy right.''